\documentclass[sigconf]{acmart}
\usepackage{algorithm}
\usepackage{algpseudocode}
\algtext*{EndWhile}
\algtext*{EndIf}
\algtext*{EndFor}
\algtext*{EndProcedure}
\usepackage{amsmath}
\usepackage{amssymb}
\usepackage{array}
\usepackage{booktabs}
\usepackage{color}
\usepackage{comment}
\usepackage{cuted}
\usepackage[inline]{enumitem}
\usepackage{epsfig}
\usepackage{etoolbox}
\usepackage{fancybox}
\usepackage{float}
\usepackage{graphicx}
\usepackage{hyperref}
\usepackage{listings}
\usepackage{lipsum}
\usepackage{multirow}
\usepackage{multicol}
\usepackage{makecell}
\usepackage{placeins}
\usepackage{pgffor}
\usepackage{rotating}
\usepackage{setspace}
\usepackage[hang, tight]{subfigure}
\usepackage{threeparttable}
\usepackage{url}
\usepackage{verbatim}
\usepackage{wrapfig}
\usepackage[dvipsnames]{xcolor}

\newbool{inccomment}
\setbool{inccomment}{true}
\graphicspath{{_fig/}}

\algrenewcommand\algorithmicindent{1.0em}%

\newcommand{\XC}[1]{\ifbool{inccomment}{{\color{magenta}YC\@: #1}}{}}
\newcommand{\ML}[1]{\ifbool{inccomment}{{\color{blue}XC\@: #1}}{}}
\newcommand{\TD}[1]{\ifbool{inccomment}{{\color{orange}#1}}{}}
\newcommand{\FN}[1]{\ifbool{inccomment}{{\color{OliveGreen}#1}}{}}

\newcommand{\ct}{\ifbool{inccomment}{{\color{magenta}$[$C$]$}}}
\newcommand{\llnn}{\ifbool{inccomment}{{\color{magenta}\\=============================================\\}}}

\newcommand{\textplaceholder}[3][gray]{%
  {%
    \color{#1}%
    \def\word{lorem}
    \if\relax\detokenize{#3}\relax\else\def\word{#3}\fi
    \foreach \i in {1,...,#2}{\word\ }%
  }%
}

\newcommand{\ours}{\textit{HeRo}}

\copyrightyear{2026}
\acmYear{2026}
\setcopyright{cc}
\setcctype{by-nc-nd}
\acmConference[DAC '26]{63rd ACM/IEEE Design Automation Conference}{July 26--29, 2026}{Long Beach, CA, USA}
\acmBooktitle{63rd ACM/IEEE Design Automation Conference (DAC '26), July 26--29, 2026, Long Beach, CA, USA}
\acmDOI{10.1145/3770743.3804310}
\acmISBN{979-8-4007-2254-7/2026/07}
\title{\ours: Adaptive Orchestration of Agentic RAG on Heterogeneous Mobile SoC}

\author{
Maoliang Li$^{1,*}$,
Jiayu Chen$^{1,*}$,
Zihao Zheng$^{1}$,
Ziqian Li$^{2}$,
Xinhao Sun$^{1}$,
Guojie Luo$^{1}$,
Chenchen Liu$^{3}$,
Xiang Chen$^{1,\dagger}$
}

\affiliation{%
  \institution{
    $^{1}$School of Computer Science, Peking University, Beijing, China\\
    $^{2}$School of Computer Science, Northwestern Polytechnical University, Xi'an, China\\
    $^{3}$School of Integrated Circuit Science and Engineering, Beihang University, Beijing, China
  }
  \country{}
}


\begin{document}

\begin{abstract}
Agentic retrieval-augmented generation (RAG) is promising for on-device deployment on mobile SoCs, but its multi-stage and dynamic workflow is difficult to schedule efficiently across heterogeneous processors due to accelerator affinity, shape sensitivity, and shared-memory bandwidth contention. 
We present \ours, a heterogeneous-aware framework for low-latency on-device agentic RAG. 
It combines profiling-based performance modeling with a lightweight online scheduler to enable shape-aware partitioning, criticality-aware mapping, and bandwidth-aware concurrency control. 
Experiments on commercial mobile devices show up to $10.94\times$ lower end-to-end latency over existing strategies, making practical on-device agentic RAG possible.
\end{abstract}

\maketitle

\begingroup
\renewcommand\thefootnote{}
\footnotetext{$^{*}$ Both authors contributed equally to this research}
\footnotetext{$^{\dagger}$ Corresponding author: Xiang Chen, xiang.chen@pku.edu.cn}
\endgroup

\section{\textbf{Introduction}}
\label{sec:introduction}

With the integration of diverse AI processing units (PUs) into mobile Systems-on-Chip(SoCs)~\cite{qualcomm8gen3} and the advancement of optimization techniques, large language model–based applications (LLM Apps) are increasingly being deployed on mobile devices.
    Given the limited computing resources, recent systems ~\cite{chen25heterollm,xu25mllmnpu} have begun exploiting concurrent execution across heterogeneous accelerators to maximize inference throughput.
    Meanwhile, because modern mobile devices store sensitive personal data that must be processed locally for privacy, on-device retrieval-augmented generation (RAG) has emerged as a highly promising scenario for mobile LLMs.

RAG augments LLMs with local context by retrieving information relevant to a user query from a vector database~\cite{douze2024faiss} and incorporating it into the prompt during response generation.
    While naive RAG adopts a simple retrieval–then–generation pipeline, recent progress in agentic RAG~\cite{tan25ayo} has substantially increased workflow complexity.
    With specialist agents such as query refiners, rerankers, and web searchers (we refer to the execution of each agent as a \textit{stage}), agentic RAG systems involve multiple LLMs with heterogeneous architectures, complex inter-stage dependencies, and dynamic runtime execution flows.

\begin{figure}[t!]
    \centering
    \includegraphics[width=3in]{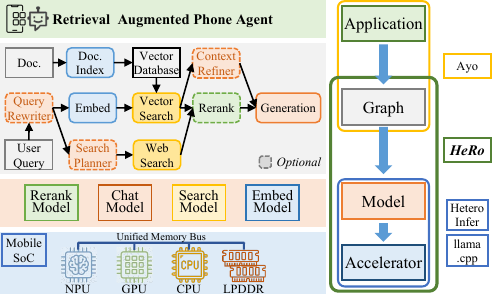}
    \vspace{-2mm}
    \caption{RAG Application Optimization Stack. Our work is a hardware and graph co-scheduling framework.}
    \Description[A]{AA}
    \label{fig:overview}
    \vspace{-8mm}
\end{figure}

Prior works on mobile LLM optimization focus primarily on model-level optimizations in the AI compiling stack illustrated in Fig.~\ref{fig:overview}.
    Existing systems either statically map a model to an accelerator (e.g., llama.cpp~\cite{llamacpp}, mllm.npu~\cite{xu25mllmnpu}) or partition operators across heterogeneous PUs (e.g., PowerInfer2~\cite{xue24powerinfer2}, HeteroInfer~\cite{chen25heterollm}).
    In contrast, agentic RAG is inherently multi-model and multi-stage: each agent stage can be decomposed into \emph{sub-stages}, forming a fine-grained task graph.
    The scheduling problem thus extends from intra-model operator placement to inter-stage coordination over a sub-stage DAG, where both graph topology and execution cost are runtime-dependent.
    This cross-layer design space from the RAG graph to the accelerator level remains largely unexplored.

\textit{From the hardware perspective}, mobile SoCs introduce three key issues.
    First, \textbf{stage--accelerator affinity}: heterogeneous PUs exhibit very different latencies for the same RAG sub-stage.
    Second, \textbf{workload shape sensitivity}: mobile runtimes and accelerators offer limited shape adaptability, making inference sensitive to input shape besides workload size.
    Third, \textbf{bandwidth contention}: all PUs share a unified DRAM; concurrent sub-stages increase traffic and can dominate the latency benefit of parallelism.

\textit{From the workflow and algorithmic perspective}, agentic RAG further complicates orchestration.
    \textbf{Intra-stage partition} is required to expose more parallelism: a logical stage can be split into sub-stages via batching or token-group partitioning, with different latency and downstream concurrency.
    At the same time, \textbf{dynamic inter-stage dependencies} cause the workflow to evolve at runtime, since agent decisions may create extra retrieval or tool calling.

These issues jointly lead to unique challenges for cross-layer scheduling:
    (1) choosing \textit{intra-stage workload partition} with proper shape and semantics,
    (2) \textit{mapping sub-stages to accelerators} with stage-PU affinity awareness,
    (3) controlling \textit{inter-stage concurrency} under shared-bandwidth interference, and
    (4) scheduling on a partial and evolving DAG with \textit{dynamic dependencies}.

To address these challenges, we propose \ours, a heterogeneity-aware RAG orchestration framework for mobile SoCs.
    \ours\ builds profiling-based performance models that capture latency and bandwidth for each model–PU combination.
    On top of this, it builds a scheduler that integrates shape-aware sub-stage partitioning, criticality-affinity joint accelerator mapping that prioritizes sub-stages on the critical path, and bandwidth-aware concurrency control that selectively enables parallel execution with minimal slowdown.
Our contributions are summarized as follows:
\begin{itemize}
    \item \textbf{Heterogeneous performance modeling.}
        We present profiling-based models for RAG stages that capture PU affinity, shape sensitivity, and bandwidth contention on mobile SoCs, enlarging the scheduling space.
    \item \textbf{Adaptive heterogeneous RAG scheduler.}
        We design an online scheduler that combines stage partition, accelerator mapping, and concurrency control for agentic RAG .
    \item \textbf{System implementation and evaluation.}
        We implement \ours\ as the first mobile system for efficient scheduling of agentic RAG workflows, and demonstrate up to $10.94\times$ latency reduction on commercial mobile phones.
\end{itemize}

\begin{table}[t]
\centering\footnotesize
\caption{Comparison with Prior Arts for RAG Deployment}
\label{tab:compare}
\vspace{-2mm}
\begin{tabular}{c|cccc|c}
\hline
                                                                 & Ayo        & HedraRAG   & llama.cpp  & HeteroInfer & Ours        \\ \hline
\begin{tabular}[c]{@{}c@{}}Hetero\\ Affinity\end{tabular}        & $\times$   & CPU-GPU    & $\times$   & GPU-NPU     & xPU         \\ \hline
\begin{tabular}[c]{@{}c@{}}Contention\\ Control\end{tabular}     & Discrete   & Discrete   & $\times$   & $\times$    & \checkmark  \\ \hline
\begin{tabular}[c]{@{}c@{}}xPU \\  Mapping\end{tabular}          & Static     & Static     & Static     & Static      & Adaptive    \\ \hline
\begin{tabular}[c]{@{}c@{}}Intra-Stage\\ Opt\end{tabular}        & $\times$   & \checkmark & Operator   & Op + Graph  & Stage       \\ \hline
\begin{tabular}[c]{@{}c@{}}Inter-Stage\\ Dependency\end{tabular} & \checkmark & \checkmark & $\times$   & $\times$    & \checkmark  \\ \hline
\end{tabular}
\vspace{-3mm}
\end{table}
\raggedbottom
\section{\textbf{Background}}
\label{sec:background}


\subsection{On-Device LLM Inference}
Considering the imperatives of personal privacy, deploying LLMs on local mobile devices is prevalent.
    Vendors are advancing their mobile platforms ~\cite{qualcomm8elite,qualcomm8gen3,mediatek9500,applea19} with various accelerators, including GPUs and neural processing units (NPUs).
To enhance the responsiveness of LLM-based apps, substantial efforts focus on fully exploiting hardware on resource-constrained mobile devices. 
    llama.cpp~\cite{llamacpp} provides highly optimized kernels and model support for diverse accelerators, establishing a foundation for LLM inference on mobile SoCs. 
    mllm.npu~\cite{xu25mllmnpu} integrates CPU/NPU co-execution with chunked prefilling and fine-grained graph scheduling, while PowerInfer2~\cite{xue24powerinfer2} employs sparse model partitioning and selective offloading. 
    Recent frameworks such as HeteroInfer~\cite{chen25heterollm} and Agent.xpu further coordinate GPU–NPU execution through fine-grained tensor partition and inter-device communication optimizations. 
    Despite these advances, existing approaches primarily target the latency optimization of a \emph{single} model and fall short in handling concurrent execution of \emph{multiple} models. 
    The unique challenges of agentic RAG systems, which require coordinated execution across multiple LLM instances with size-varying workloads and data dependencies, are still unaddressed.

\begin{table}[b!]
\centering\small
\setlength{\tabcolsep}{13pt}
\caption{Specifications of Mobile SoCs Adopted in this Work}
\label{tab:socspec}
\vspace{-2mm}
\begin{tabular}{@{}ll@{}}
\toprule
\multicolumn{2}{c}{\textbf{Redmi K80 (Snapdragon 8 Gen 3 SoC)}}                 \\ \midrule
\multicolumn{1}{l|}{CPU}    & Kryo, 8-core, up to 3.4 GHz                       \\
\multicolumn{1}{l|}{GPU}    & Adreno 750 | FP16=2.8 TFlops                      \\
\multicolumn{1}{l|}{NPU}    & Hexagon v75 | INT8=34Tops                         \\
\multicolumn{1}{l|}{Memory} & 12GB LPDDR5x | Bandwidth: 76.8 GB/s with 64 bits  \\ \midrule

\multicolumn{2}{c}{\textbf{OnePlus 13 (Snapdragon 8 Gen 4 SoC)}}                \\ \midrule
\multicolumn{1}{l|}{CPU}    & Oryon, 8-core, up to 4.5 GHz                      \\
\multicolumn{1}{l|}{GPU}    & Adreno 830 | FP16=3.4 TFlops                      \\
\multicolumn{1}{l|}{NPU}    & Hexagon v79 |  INT8=50Tops                        \\
\multicolumn{1}{l|}{Memory} & 24GB LPDDR5x | Bandwidth: 84.8 GB/s with 64 bits  \\ \bottomrule
\end{tabular}
\vspace{-2mm}
\end{table}

\subsection{Agentic RAG Workflows}

In general, RAG systems follow a retrieval-then-generation pipeline. 
    However, this structure struggles with complex inputs containing ambiguous or multifaceted queries. 
    To enhance retrieval accuracy, recent works incorporate specialist agents~\cite{wu2023autogen,gao23hydeprecise,xu2024recomp}.
    Pre-processing agents, such as query rewriters and decomposers, transform casual user inputs into search-friendly queries or structured sub-queries. 
    Conversely, post-processing agents, including document rerankers and context refiners, aim to provide more relevant and coherent context to the LLM, thereby improving generation quality. 
    These optimizations are particularly beneficial for on-device LLMs with limited generalist knowledge.
    Nevertheless, the introduction of multiple agents brings significant multi-model coordination~\cite{qwen3embedding,qwen3technicalreport} and increases runtime dynamism, posing new challenges for efficient execution on mobile platforms.

\subsection{RAG Deployment Optimization}
With the prevalence of RAG, deployment optimization becomes a pivotal problem.
    Most works lie in \textit{stage-specific optimization}, such as long-context LLM generation~\cite{xu25mllmnpu}, vector search ~\cite{park2025mobilerag} and reranking~\cite{zhou2025grating}.
    In contrast, workflow-level optimization orchestrating multiple stages for higher throughput emerges in cloud services like \cite{jiang25rago,yu25ragdoll}.
    On the edge side, Ayo~\cite{tan25ayo} and HedraRAG~\cite{hu25hedrarag} make preliminary progress. Ayo attempts to improve intra-stage parallelism through task decomposition, but suffers from redundant computation and limited adaptability under dynamic loads. 
    HedraRAG addresses interleaved retrieval–generation scheduling for hybrid CPU-GPU cooperation, yet its optimization remains restricted to a narrow set of model types and lacks system-level heterogeneity awareness.
    When considering on-device scenarios with complex agentic workflows, there is still a gap in efficient mapping from workflow stages to multiple accelerators without exploiting hardware characteristics.
    
\section{\textbf{Agentic RAG on Heterogeneous SoCs}}
\label{sec:analysis}

In this section, we provide analysis and modeling for agentic RAG workflow inference on heterogeneous accelerators. In this work, for development convenience, we use Qualcomm's Snapdragon 8Gen3 and 8~Elite (8Gen4)~\cite{qualcomm8elite,qualcomm8gen3} as specified in Tab~\ref{tab:socspec}.

\subsection{Agentic RAG Workflows}

As illustrated in Fig.~\ref{fig:overview}, agentic RAG introduces substantially more complex workflows than classical two-stage retrieval–generation pipelines. We identify two key characteristics.

\textbf{Intra-Stage Partition.}
Although the user application specifies the high-level workflow between agents, a given agent may be instantiated multiple times during execution.
    Moreover, stage-level dependencies often introduce unnecessary synchronization. 
    For example, a query rewriter may issue several search requests, but the first search need not wait for all subsequent queries to be generated.
    To expose additional parallelism, we model dependencies at a finer granularity by decomposing stages into \emph{sub-stages}, which represent a partition of the workload within a stage, such as indexing part of the documents or decoding a sub-query.

\textbf{Dynamic Inter-Stage Dependencies.}
In agentic RAG, the execution flow is not predetermined, but rather evolves based on the responses of the agents.
    Agents may generate responses of varying lengths, perform additional retrieval steps, or skip optional stages. 
    This means that the graph is partially observable.
    At time \(t\), the scheduler observes only a \emph{partial} DAG \(G^{\text{obs}}(t)\), where \(V^{\text{obs}}(t) \subseteq V\) and \(E^{\text{obs}}(t) \subseteq E\).
    New nodes and edges materialize as upstream decision stages finish.
    This dynamic depth requires scheduling under incomplete future knowledge, leveraging statistical priors over typical agent behaviors to anticipate downstream computation.
\begin{figure}[t]
    \centering
    \includegraphics[width=3.3in]{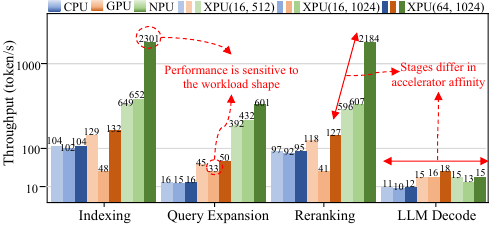}
    \vspace{-2mm}
    \caption{Stage-Accelerator Affinity and Shape Sensitivity.}
    \Description[A]{AA}
    \label{fig:analysis-shape}
    \vspace{-2mm}
\end{figure}

\subsection{Modeling RAG on Heterogeneous SoCs }
We consider an SoC equipped with a set of heterogeneous processing units (PUs) \(\mathcal{K}\), which share a unified DRAM subsystem.
Under this setting, RAG inference exhibits the following characteristics.

\textbf{Stage-Accelerator Affinity. }
For each sub-stage $v \in V$ and PU $k \in \mathcal{K}$, differences in model implementations and kernel optimizations yield distinct behaviors.
    We capture this via a configuration set \(\mathcal{C}_v\), where each configuration $c \in \mathcal{C}_v$ specifies the target PU $k_c$ and a workload shape $s_v(c)$ (e.g., batch size or sequence length).
    Given $c$, the \emph{base latency} $p_v^0(c)$ without interference is profiled offline.
    The affinity between stages and accelerators is reflected in the variation among base latencies.
    For example, stages such as indexing or reranking often run much faster on NPUs, whereas LLM generation stages favor GPUs, as shown in Fig.~\ref{fig:analysis-shape}.

\textbf{Workload Shape Sensitivity.}
Many stages, such as document indexing or reranking, can naturally process multiple items in batches.
    For a fixed stage and PU, the base execution time depends on batch size in a highly non-linear manner due to effects such as tiling, kernel fusion, and memory hierarchy behavior.
    Formally, \(p_v^0(c) = f_{m,k}\big(s_v(c)\big)\), where $f$ is obtained via offline profiling.
    Larger batches do not always yield better per-item efficiency, as shown in Fig.~\ref{fig:analysis-shape}.
    Thus, the orchestrator must selectively exploit batch-size sensitivity for batchable stages, while relying on sub-stage partitioning for streaming stages, i.e., autoregressive decoding.

\begin{figure}[t]
    \centering
    \includegraphics[width=3.3in]{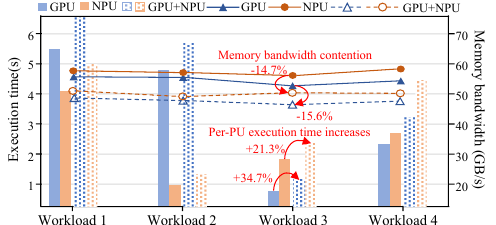}
    \vspace{-2mm}
    \caption{Contention Slowdown under Various Parallelism.}
    \Description[A]{AA}
    \label{fig:analysis-bandwidth}
    \vspace{-2mm}
\end{figure}

\textbf{Inter-Stage Bandwidth Contention.} 
On mobile SoCs, all accelerators share a unified DRAM subsystem with peak bandwidth $B_0$. 
    When multiple independent stages execute on different PUs, bandwidth contention may incur latency overhead, as depicted in Fig.~\ref{fig:analysis-bandwidth}.
    Each configuration $c$ of a sub-stage $v$ incurs a bandwidth demand $b_v(c)$, estimated via profiling-based modeling. 
    At time $t$, let $\mathcal{A}(t)$ denote the set of active sub-stages; the aggregate bandwidth demand is \(B(t) = \sum_{v \in \mathcal{A}(t)} b_v(c_v)\).
    Rather than treating $B_0$ as a hard limit, we model contention-induced slowdown using a factor $\phi_v(B(t)) \ge 1$ that increases monotonically with $B(t)$:
        when $B(t)\!\ll\! B_0$, $\phi_v(B(t)) \approx 1$; as $B(t)$ approaches or exceeds $B_0$, $\phi_v(B(t))$ grows due to rising DRAM latency and reduced effective throughput.
    The function $\phi_v(\cdot)$ is stage- and configuration-dependent, capturing model-level and hardware-level sensitivity to bandwidth pressure.
    The effective execution time can now be approximated as:
\begin{equation}
    p_v(c_v, B(t)) \approx p_v^0(c_v)\, \phi_{v,c}(B(t)).
\end{equation}

\begin{figure*}[tb]
    \centering
    \includegraphics[width=7.2in]{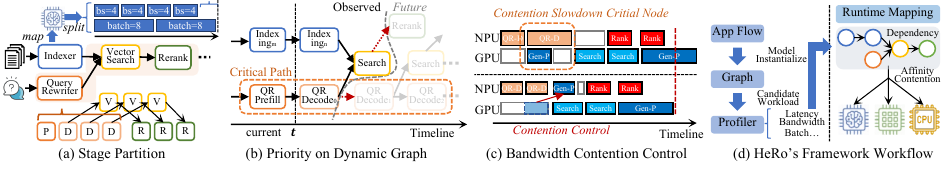}
    \vspace{-8mm}
    \caption{Orchestration Techniques. }
    \Description[A]{AA}
    \label{fig:methods}
    \vspace{-2mm}
\end{figure*}

\subsection{Heterogeneous Deployment Challenges}

The combination of agentic RAG workflows and heterogeneous mobile SoCs introduces several deployment challenges:

\noindent\textbf{Intra-Stage Workload Partitioning.}
    Many stages support multiple workload shapes and partitioning strategies, leading to various per-PU latencies and degrees of downstream parallelism.

\noindent\textbf{Stage–Accelerator Mapping.}
    Each stage exhibits heterogeneous performance across PUs.  
    The scheduler must assign stages to accelerators to exploit affinity while avoiding workload imbalance.

\noindent\textbf{Inter-Stage Concurrency Control.}
    Although concurrent use of multiple accelerators improves utilization, for a latency-critical request, excessive parallelism can degrade performance when bandwidth contention induces slowdown.

\noindent\textbf{Scheduling with Dynamic Dependency.}
    The workflow graph evolves based on agent decisions.  
    The scheduler must operate on a partial and evolving DAG, using statistical priors to anticipate and adapt to future stages.

\section{\textbf{Orchestration Methods}}
\label{sec:methods}

\subsection{Problem Formulation}
We focus on minimizing the latency of a single RAG graph, with each node $v\in V$ being a sub-stage associated with:
    (1) model,
    (2) configuration set including the PU and workload shape.
    and (3) profiled performance model for runtime, bandwidth, and slowdown.
For each $v$, the scheduler chooses a configuration $c_v \in \mathcal{C}_v$ and a start time $S_v \ge 0$.
The end-to-end latency of the RAG flow is:
\begin{equation}
  \label{equ:opt_obj}
  T = \max_{v \in V} F_v, \text{where } F_v = S_v + p^0_v(c_v)\cdot \bar{\phi}_v,
\end{equation}
where \(F_v\) denotes the completion time of $v$, $\bar{\phi}_v$ is an average slowdown factor over the execution of $v$, induced by the time-varying $B(t)$ and $\phi_v(\cdot)$.
Minimizing $T$ requires determining $\{c_v, S_v\}_{v \in V}$ under dependency and interference, forming a variant of the precedence-constrained task-graph scheduling problem on unrelated machines which is NP-hard~\cite{drozdowski09scheduling}.
    We therefore design online heuristics that exploit the specific structure of the agentic RAG workflow and the slowdown model.


\subsection{Adaptive Heterogeneous Scheduling}

Our orchestration algorithm consists of three key components, namely the sub-stage partitioner, the priority estimator, and the concurrency controller, combined into an online heterogeneous scheduler for effectively handling RAG execution challenges.

\textbf{Shape-Aware Sub-Stage Partition.}
Proper sub-stage boundaries and batching decisions are crucial for balancing latency and execution efficiency, as illustrated in Fig.~\ref{fig:methods}(a).
    For streaming workloads such as query rewriting, scheduling is performed at the granularity of token groups to amortize scheduling overhead. 
        As decoding progresses, the runtime monitors the generated content and triggers downstream stages once their data dependencies are satisfied.
    For batchable stages (e.g., indexing document chunks), we offline profile a candidate set of batch sizes $\mathcal{N}_{m,k}$ for each model-PU pair $(m,k)$, recording both latency and bandwidth usage. 
        Given an input workload, the runtime partitions the stage by selecting the batch size that minimizes the estimated execution time. 
        We assume each stage executes on a single PU; if inter-PU migration occurs, partitioning is recomputed based on the remaining workload.
        Specifically, for stage $m$ with workload size $L$ executed on PU~$k$, we search for the optimal workload unit corresponding to batch size $n$:
\begin{equation}
    \label{equ:partition}
    \arg\min_{n \in \mathcal{N}_{m,k}} \left\lceil L/n \right\rceil \cdot p_{v}^0(c_v), \text{where } v=m,c_v=(n, k).
\end{equation}

\textbf{Critical-Stage Prioritized Accelerator Mapping.}
On-device RAG inference is single-batched. Thus, mapping stages on the critical path to the fastest PU is pivotal to latency reduction.
Since the full execution graph is unknown at runtime (Fig.~\ref{fig:methods}(b)), we estimate node priority using a \textit{criticality score} composed of an \emph{observed} term and a \emph{future} term. 
    At time $t$, the runtime maintains $G^{\text{obs}}(t)$ and the \emph{ready set} $\mathcal{R}(t)$ containing nodes whose predecessors have completed.
    The observed term $CS_{\text{L}}(v)$ is computed on $G^{\text{obs}}(t)$ as the longest remaining path from $v$ to any undetermined node. 
        During this simulation, PU assignment follows a dependency-agnostic heuristic similar to SJF, prioritizing shorter-latency tasks; $CS_{\text{L}}(v)$ is updated whenever $G^{\text{obs}}(t)$ evolves.
    To account for uncertain future execution, we compute the future term $CS_{\text{F}}(v)$ on a predefined RAG workflow graph, using historical averages to estimate the likelihood of downstream activation. 
        Intuitively, agents such as the search planner tend to trigger more subsequent computation than lightweight post-processing modules, resulting in higher expected future criticality.
    The final criticality score is defined in Eq.~\ref{equ:criti_score}, where $\beta$ controls the influence of future dependencies. 
        Nodes with larger $CS(v)$ are prioritized, as they lie on long observed paths and are more likely to unlock future work:
\begin{equation}
\label{equ:criti_score}
    CS(v) = CS_{\text{L}}(v) + \beta \cdot CS_{\text{F}}(v).
\end{equation}
    
\textbf{Bandwidth-Aware Concurrency Control.}
Under single-batch execution and shared DRAM, overly aggressive parallelism can stall nodes on the critical path and increase end-to-end latency, as exemplified in Fig.~\ref{fig:methods}(c). 
    To avoid this, we limit concurrency to prevent harmful interference, even when doing so temporarily reduces PU utilization.
    When PU~$k$ becomes idle, the scheduler considers issuing a ready node $v \in \mathcal{R}(t)$ with configuration $c_v$. 
    The slowdown incurred by admitting $v$ is estimated as $\phi_v\!\left(B(t) + b_v(c_v)\right)$. 
    We first enforce a soft bandwidth constraint \(B_{\text{soft}} \) to avoid system-wide degradation.
    Because evaluating global performance impact is costly, we approximate it by measuring interference on the current critical-path node $v^*$, defined as the ready or running node with the highest criticality score $CS(v^*)$.
    We introduce a contention penalty in Eq.~\ref{equ:bwpanelty} that jointly reflects contention-induced slowdown and node criticality, where $(t - S_{v^*})$ is the time $v^*$ has been active.
    This formulation admits parallelism only when it does not significantly impede critical-path progress, favoring decisions that minimize end-to-end latency rather than maximize instantaneous PU utilization:
\begin{equation}
\label{equ:bwpanelty}
    W_B = 
    \phi_{v^*}\!\big(B(t) + b_v(c_v)\big)\,
    (t - S_{v^*})\,
    p_{v^*}(c_{v^*}).
\end{equation}

\textbf{Overall Scheduler.}
Alg.~\ref{alg:scheduler} outlines the scheduler of \ours:
    At each step, the scheduler identifies the most critical ready node $v^{\text{cand}}$ according to Eq.~\ref{equ:criti_score}, and selects the PU and configuration that minimize the predicted completion time with contention penalty.
    For all PUs capable of executing $v^{\text{cand}}$, the runtime enumerates a small set of shape-aware configurations and prunes those that violate the bandwidth budget.
    For each remaining configuration, it estimates the execution time of $v^{\text{cand}}$ and its interference on the current critical-path node $v^*$ using the contention penalty $W_B(v^*, v^{\text{cand}}, t, c)$, then chooses the configuration with the lowest overall score.
    If there is no feasible configuration for the current most critical node, the scheduler defers it and considers the second critical node.
This ensures that high-criticality nodes are dispatched to PUs that complete them earliest, while still respecting concurrency constraints.

\begin{algorithm}[tb]
\footnotesize
\caption{\small Node-centric Online Heterogeneous RAG Scheduler}
\label{alg:scheduler}
\begin{algorithmic}[1]
\State \textbf{Input:} profiling model $p^0_v(c), b_v(c), \phi_v(\cdot)$, bandwidth budget $B_{\text{soft}}$
\While{there exists an unfinished node in $V$}
  \State Update $G^{\text{obs}}(t)$, $\mathcal{R}(t)$ with finished nodes 
  \State Update criticality ${CS}(v)$ for $v \in \mathcal{R}(t) \cup \mathcal{A}(t)$ \Comment{Eq.~\ref{equ:criti_score}}
  \State $\mathcal{R}_{\text{tmp}} \leftarrow \mathcal{R}(t)$
  \While{exists idle PU and $\mathcal{R}_{\text{tmp}} \neq \emptyset$}
    \State $v^* \leftarrow \arg\max_{x \in \mathcal{R}(t) \cup \mathcal{A}(t)} {CS}(x)$ \Comment{current critical-path node}
    \State $v^{\text{cand}} \leftarrow \arg\max_{v \in \mathcal{R}_{\text{tmp}}} {CS}(v)$ \Comment{most critical ready node}
    \ForAll{PU $k$ that can execute $v^{\text{cand}}$}
      \ForAll{$c \in \text{ShapeAwareConfigs}(v^{\text{cand}},k)$} \Comment{Eq.\ref{equ:partition}}
        \If{$B(t) + b_{v^{\text{cand}}}(c) > B_{\text{soft}}$} \textbf{continue} \EndIf
        \State Calculate $F_{v^{\text{cand}}}(c)$ and $W_B(v^*,v^{\text{cand}}, t,c)$ with Eq~\ref{equ:opt_obj} and Eq~\ref{equ:bwpanelty}
        \State $\text{score}(c) \leftarrow F_{v^*}(c) + \alpha \cdot W_B(v^*,v^{\text{cand}}, t,c)$
        \State $\mathcal{C}_{v^{\text{cand}}}^{\text{all}} \leftarrow \mathcal{C}_{v^{\text{cand}}}^{\text{all}} \cup \{(k,c)\}$
      \EndFor
    \EndFor
    \If{$\mathcal{C}_{v^{\text{cand}}}^{\text{all}} = \emptyset$} 
      $\mathcal{R}_{\text{tmp}} \setminus \{v^{\text{cand}}\}$, \textbf{continue} 
      \Comment{try next critical node}
    \EndIf
    \State $c_\text{cand} \leftarrow \arg\min \text{score}(c)$, where $c\in \mathcal{C}_{v^\text{cand}}^{\text{all}}$
    \State Dispatch $v^{\text{cand}}$ on PU $k^*$ with config $c^{\text{cand}}$; $S_{v^{\text{cand}}} \leftarrow t$
    \State  $\mathcal{R}(t) \leftarrow \mathcal{R}(t)\setminus \{v^{\text{cand}}\}$, 
            $\mathcal{R}_{\text{tmp}} \leftarrow \mathcal{R}_{\text{tmp}} \setminus \{v^*\}$
    \State  $\mathcal{A}(t) \leftarrow \mathcal{A}(t)\cup \{v^*\}$,
            $B(t) \leftarrow B(t) + b_{v^{\text{cand}}}(c^{\text{cand}})$
  \EndWhile
  \State \textbf{wait} until any $v\in\mathcal{A}(t)$ finishes; Update $\mathcal{A}(t), B(t)$ and current time $t$
\EndWhile
\end{algorithmic}
\end{algorithm}

\section{\textbf{Implementation}}
\label{sec:impl}

We implemented \ours~ on top of llama.cpp~\cite{llamacpp}, Powerserve~\cite{powerserve} (a mobile LLM inference framework with Hexagon NPU support), and Faiss~\cite{douze2024faiss}, with roughly 5{,}000 lines of C/C++ code.
    \ours\ reuses the CPU and OpenCL backends from llama.cpp and integrates the NPU backend from Powerserve. 
    We extend the runtime to support both embedding and reranking model inference, as well as chunked prefill mechanisms. 
    To enable efficient heterogeneous execution, we establish shared memory between CPU and GPU via OpenCL, and between CPU and NPU using the QNN API~\cite{qualcomm2025qnn}. 
    Furthermore, by leveraging the \texttt{CL\_MEM\_USE\_HOST\_PTR} flag, we map NPU-side buffers into the GPU address space, enabling zero-copy data sharing among all three processors.
    Following prior online scheduling work~\cite{jia22CoDL,jeong22Band}, we estimate latency, bandwidth requirements, and slowdown functions for irregular workload sizes using a multi-feature linear regression model trained on sampled measurements across different models, workload sizes, and background contention levels.
    The scheduler exposes two hyperparameters: the bandwidth-contention penalty weight $\alpha$ in priority assignment and the future-term weight $\beta$ in criticality estimation.
    We tune both parameters for each deployment via grid search.


\section{\textbf{Experiments}}
\label{sec:experiments}

\subsection{Experiment Setup}

\noindent\textbf{Applications.} 
Our experiments cover three agentic RAG workflows with varying complexity, derived from the template workflow in Fig.~\ref{fig:overview}.
    \textbf{\textit{Workflow~1}}, \emph{Fast Document Finder}, segments documents into chunks (default size: 128, overlap: 10), embeds them, and stores them in the vector DB. Then it retrieves and reranks the most relevant chunks to generate the final response.
    \textbf{\textit{Workflow~2}}, \emph{Advanced Document QA Bot}, extends Workflow~1 by enabling LLM-based query rewriting and context refinement, trading off latency for improved retrieval and comprehension accuracy.
    \textbf{\textit{Workflow~3}}, \emph{Deep Researcher}, further incorporates online resources, with a lightweight LLM–based search planner for issuing web-search requests.
    
\noindent\textbf{Datasets.}
Following prior work~\cite{tan25ayo,hu25hedrarag} on RAG workflow scheduling, we adopt four widely used RAG datasets with diverse workload characteristics: FinqaBench~\cite{islam2023finqabench}, TruthfulQA~\cite{lin2022truthfulqa}, HotpotQA~\cite{yang2018hotpotqa}, and 2WikiMultihopQA~\cite{xanh202wikimultihop}. 
    FinqaBench and TruthfulQA generally contain shorter inputs (query length $\le$ 70 tokens; context $\sim$200 tokens), whereas HotpotQA and 2WikiMultihopQA include longer and more complex contexts, reaching up to 1k tokens. 
    To accommodate mobile devices' limited computing capability, we exclude datapoints with extremely long queries or contexts.

\begin{figure}[t]
    \centering
    \includegraphics[width=3.3in]{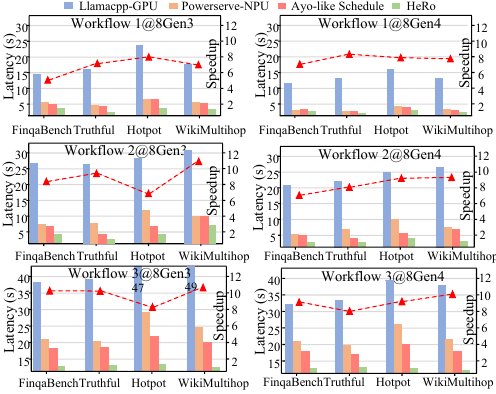}
    \vspace{-2mm}
    \caption{End-to-End Latency on Qwen3 Family. Embed: Qwen3-Embedding-0.6B, Rerank : Qwen3-Reranker-0.6B, Search: Qwen3-1.7B, Chat: Qwen3-4B.}
    \Description[A]{AA}
    \label{fig:result_cfg1}
    \vspace{-2mm}
\end{figure}

\begin{figure}[t]
    \centering
    \includegraphics[width=3.3in]{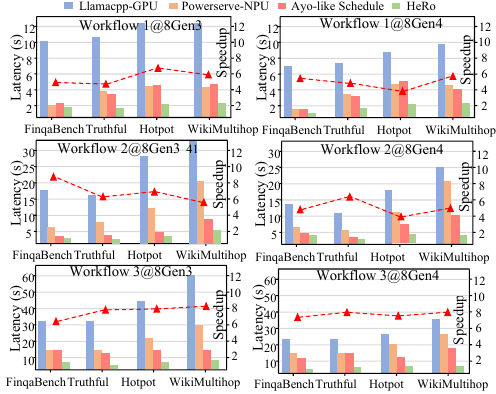}
    \vspace{-2mm}
    \caption{End-to-End Latency on BGE and LLaMA3 Family. Embed: bge-large-en-v1.5(0.3B), Rerank: bge-reranker-large(0.6B), Search: Llama-3.2-1B, Chat: LLaMA-3.1-8B.}
    \Description[A]{AA}
    \label{fig:result_cfg2}
    \vspace{-2mm}
\end{figure}

\noindent\textbf{Models.}
To evaluate the scalability of \ours, we instantiate the workflows using two families of on-device LLMs with different model scales and architectures: 
    (1) the Qwen3 series~\cite{qwen3embedding,qwen3technicalreport}, and 
    (2) the bge series~\cite{chen2024bge-emb,li2023bge-rerank} together with Llama3~\cite{grattafiori2024llama3herdmodels}. 
    The detailed workflow–model compositions are shown in Fig.~\ref{fig:result_cfg1} and Fig.~\ref{fig:result_cfg2}, where all models are quantized to INT8.
    
\noindent\textbf{Baselines.}
To our knowledge, few studies have specifically optimized RAG, especially agentic RAG workflows, on heterogeneous mobile SoCs. 
    By composing RAG workflows on top of popular mobile LLM frameworks, we construct three strong baselines:
    (1) \textbf{\textit{Llama.cpp-GPU}}: all models except FAISS run on GPU with the OpenCL backend adapted from \texttt{llama.cpp}; 
    (2) \textbf{P\textit{owerserve-NPU}}: all models except FAISS run on NPU with the QNN backend adapted from Powerserve; 
    (3) \textbf{\textit{Ayo-like}}: models are manually mapped to all xPUs based on workflow dependencies and model sizes.

\noindent\textbf{Metric.}
Mobile users typically issue a single request at a time. 
    Thus, we mainly evaluate \ours\ using average end-to-end latency for each query. 
    For fair comparison, we remove traces with abnormally short or long latencies, which often arise from unexpected behaviors due to the limited capability of on-device LLMs.

%

\subsection{Evaluation Results}

\noindent\textbf{End-to-End Speedup.} 
\ours\ delivers consistent improvements over all baselines across datasets and workflows. 
\texttt{llama.cpp} shows substantially higher latency due to the limited compute capability of mobile GPUs. 
	Although Powerserve provides noticeably better performance, it still suffers from under-utilized resources, making multi-accelerator placement inherently more effective. 
	However, static mappings cannot address the utilization loss caused by dependency constraints and heterogeneous workload affinity. 
	By performing online stage-accelerator mapping with heterogeneity-aware affinity modeling, \ours\ achieves up to $1.5\times$ speedup over the Ayo-like baseline and up to 10.94$\times$ improvement over the GPU-only execution. 
	These results highlight the necessity of exploiting multiple accelerators jointly and performing affinity-aware scheduling.

\noindent\textbf{Workflow-level Analysis.} Workflow~1 is relatively simple and exposes limited parallelism, so \ours\ yields modest gains over baseline. 
	Most of the execution time is dominated by the document-indexing stage, which is already highly NPU-friendly; hence, longer documents further diminish the attainable speedup. 
	In contrast, Workflow~3 contains the most complex execution graph. 
		Single-processor baselines perform substantially worse due to strictly sequential execution. 
\ours\ achieves the highest acceleration with minimal PU under-utilization. 
	Even compared to the Ayo-like strategy, \ours\ still provides a notable speedup because the static Ayo-like mapping often stalls on dependency bottlenecks and delays the execution of critical-path stages.

\noindent\textbf{Dataset-level Analysis.} The runtime of linear baselines scales nearly proportionally with input size; datasets with long contexts, such as HotpotQA, incur much larger slowdowns. 
    For \ours, the observed speedup is less tightly correlated with input length, as the agent may trigger additional computation when processing difficult queries—particularly when the query rewriter or search planner is activated in the more complex Workflows~2 and~3.

\begin{table}[b!]
\centering\small
\setlength{\tabcolsep}{1.5pt}
\caption{SpeedUp Breakdown of Proposed Techniques}
\label{tab:ablation}
\vspace{-4mm}
\begin{tabular}{c|cccc}
\hline
Technique              & $C_1$Latency(s) & $C_1$SpdUp  & $C_2$Latency(s)  & $C_2$SpdUp\\ \hline
Baseline               &   5.79s         &  1.0x       &    17.23s        &   1.0x     \\
+ Sub-Stage Partition  &   5.07s         &  1.14x      &    8.79s         &   1.96x     \\
+ Criticality Guidance &   4.23s         &  1.37x      &    6.81s         &   2.53x     \\
+ Concurrency Control  &   4.63s         &  1.25x      &    8.24s         &   2.09x     \\ \hline
ALL                    &   3.82s         &  1.52x      &    5.38s         &   3.20x     \\ \hline
\end{tabular}
\end{table}

\noindent\textbf{Model-level Analysis.} 
\ours\ provides larger relative speedups when using the Qwen3 family. 
    This is primarily because the Qwen3 configuration employs a smaller 4B chat model, whereas the alternative configuration uses an 8B Llama~3.1 model. 
    The smaller model results in a more balanced workload distribution and reduces hard dependencies that cannot be mitigated by sub-stage partitioning. 
    This effect is especially pronounced in Workflow~1, where sequential components dominate overall latency.

\noindent\textbf{Platform-level Analysis.} 
Although \ours\ consistently accelerates execution on both platforms with different compute capabilities, the degree of speedup varies. 
    This variation stems from the discrepancy between NPU performance and the memory-bandwidth-to-compute ratio across platforms. 
    For instance, on the 8~Gen~3 platform, larger relative speedups are observed for Workflows~2 and~3 because the memory bandwidth is higher relative to the NPU FLOPs, enabling more effective concurrent accelerator utilization.

\noindent\textbf{Ablation Studies.}
We further explore how the components of our method contribute to the result.
    Performance was measured using two samples on the 8 Gen 4 platform: 
        $C_1$: Qwen3 family configuration with Workflow 2 and input data sampled from FinqaBench. 
        $C_2$: BGE family configuration with Workflow 3 and input data sampled from 2WikiQA.
    The baseline is ``Ayo-like'' static mapping, and the results are illustrated in Tab~\ref{tab:ablation}.

\section{\textbf{Conclusion}}
\label{sec:conclusion}

This paper presents \ours, an adaptive orchestration system designed to address the challenges of deploying complex agentic RAG workflows onto heterogeneous mobile SoCs, including stage-accelerator mapping, workload partitioning, concurrency control, and dynamic dependency scheduling.
By designing a dependency-guided scheduler with awareness of shape, bandwidth, and hardware affinity, \ours\ achieves efficient resource utilization and reduced latency, achieving up to $10.94\times$ speedup.
Experimental results highlight \ours's effectiveness in optimizing RAG inference and its potential for broader agentic applications on personal devices in real-world mobile usage scenarios.

\begin{acks}
This work is supported by project No.ZR2024LZH001 supported by Shandong Provincial Natural Science Foundation.
\end{acks}

\clearpage
\balance
\bibliographystyle{ACM-Reference-Format}
\bibliography{
    _ref/1_rag_accel,
    _ref/2_dataset,
    _ref/3_related,
    _ref/98_framework_tool,
    _ref/mobile_llm,
    _ref/hardware,
    _ref/misc,
    _ref/multi_dnn,
    _ref/LLM_models
}
\end{document}